\documentclass{PoS_new}

\usepackage{amsfonts,amsmath,amssymb}
\usepackage{graphicx}
\usepackage{mathrsfs}
\usepackage{slashed}

\newcommand{\nb}{\phantom{0}}

\newcommand{\ol}[1]{\overline{#1}}

\title{Axial couplings of heavy hadrons from domain-wall lattice QCD}

\ShortTitle{Axial couplings of heavy hadrons from domain-wall lattice QCD}

\author{William Detmold$^{a,b}$, C.-J. David Lin$^{c,d}$, \speaker{Stefan Meinel}$\:^a$\\ \\
\llap{$^a$}Department of Physics, College of William \&  Mary, Williamsburg,
  VA 23187-8795, USA\\
\llap{$^b$}Jefferson Laboratory, 12000 Jefferson Avenue, Newport News, VA 23606, USA\\
\llap{$^c$}Institute of Physics, National Chiao-Tung University, 
Hsinchu 300, Taiwan\\
\llap{$^d$}Physics Division, National Centre for Theoretical Sciences,
Hsinchu 300, Taiwan\\
\\ \\
E-mail: \email{smeinel@wm.edu}}

\abstract{We calculate matrix elements of the axial current for static-light mesons and baryons
in lattice QCD with dynamical domain wall fermions. We use partially quenched heavy hadron
chiral perturbation theory in a finite volume to extract the axial couplings $g_1$, $g_2$, and $g_3$
from the data. These axial couplings allow the prediction of strong decay rates and enter chiral extrapolations
of most lattice results in the $b$ sector. Our calculations are performed with two lattice spacings
and with pion masses down to 227 MeV.}

\FullConference{ The XXIX International Symposium on Lattice Field Theory - Lattice 2011\\
July 10-16, 2011\\
Squaw Valley, Lake Tahoe, California}

\begin{document}

\section{Introduction}

The low-energy dynamics of heavy-light mesons and baryons can be described by
heavy-hadron chiral perturbation theory (HH$\chi$PT), an effective field theory for QCD
that incorporates both chiral symmetry and heavy-quark symmetry \cite{Wise:1992hn}.
HH$\chi$PT is essential for controlling light-quark-mass extrapolations of lattice QCD data in the heavy-quark sector (see for example Ref.~\cite{Kronfeld:2002ab}).

At leading order in the heavy-quark and chiral expansions, the HH$\chi$PT Lagrangian contains three axial coupling constants that determine the strength
of the interactions between heavy-light hadrons and pions: one coupling (denoted as $g_1$) for the heavy-light mesons, and
two additional couplings (denoted as $g_2$, $g_3$) for the heavy-light baryons. These axial couplings
are calculable from QCD, and their determination enables quantitative predictions
for many heavy-light hadron properties (such as masses, decay widths, and various matrix elements) using HH$\chi$PT. The chiral loop contributions
that lead to the nonanalytic dependence of such properties on the light-quark masses are proportional to
products of the relevant axial couplings. While $g_1$ has received much attention in the past because
of its role for $B$ mesons, the lesser-known couplings $g_2$ and $g_3$ are important for flavor physics with heavy baryons.
The bottom baryon sector provides complementary information to $B$ mesons for constraining the helicity structure of possible new physics \cite{Mannel:1997xy}.

The calculation of $g_{1,\:2,\:3}$ from the underlying theory of QCD must be done nonperturbatively, and hence on a lattice.
The mesonic coupling $g_1$ had been studied previously in lattice QCD with $n_f=0$ or $n_f=2$ dynamical flavors
\cite{deDivitiis:1998kj}.
In the following, we present a complete determination of all three axial couplings $g_{1,\:2,\:3}$
using $n_f=2+1$ domain-wall lattice QCD \cite{Detmold:2011bp, g123long}.
Our choice of lattice parameters (low pion masses, large volume, two lattice spacings) and
our analysis method (fits to the axial-current matrix elements using the correct next-to-leading-order
formulae from HH$\chi$PT \cite{Detmold:2011rb}) allow us to control all sources of systematic uncertainties.

\section{Heavy-hadron chiral perturbation theory}

We begin with an introduction to HH$\chi$PT. This theory combines the chiral expansion
with an expansion in powers of $\Lambda_{QCD}/m_Q$, where $m_Q$ is the heavy-quark mass.
We work at the leading order in the heavy-quark expansion, where the spin of the light degrees of freedom
($s_l$) is conserved and the heavy-quark spin decouples. The lowest-lying heavy-light
mesons with $s_l=1/2$ form multiplets with $J=0$ and $J=1$, which can be combined into a single field $H^i$:
\begin{equation}
 H^i=\left[ - P^i\gamma_5 + P^{*i}_{\mu} \gamma^\mu \right]\frac{1-\slashed{v}}{2},\hspace{1.5ex}\mathrm{with}\hspace{1.5ex} (P^i)=\left( \begin{array}{c} B^+ \\ B^0 \end{array} \right), \hspace{1ex} (P^{*i})_\mu=\left( \begin{array}{c} B^{*+} \\ B^{*0} \end{array} \right)_{\hspace{-1ex}\mu}. \label{eq:Hfield}
\end{equation}
(We consider $SU(2)$ HH$\chi$PT and use the notation for bottom hadrons.) Similarly, the baryons with $s_l=1$
form multiplets with $J=1/2$ and $J=3/2$. These are described by Dirac and Rarita-Schwinger fields $B^{ij}$ and $B^{*ij}_\mu$, which
are symmetric in the flavor indices and can be combined into a single field $S^{ij}$:
\begin{eqnarray}
\nonumber S_\mu^{ij}&=&\sqrt{\frac{1}{3}}(\gamma_\mu + v_\mu)\gamma_5 B^{ij} + B^{*ij}_{\mu},\hspace{1.5ex}\mathrm{with}\hspace{1.5ex}(B^{ij})=\left( \begin{array}{cc} \Sigma_b^+ & \frac{1}{\sqrt{2}}\Sigma_b^0 \\ \frac{1}{\sqrt{2}}\Sigma_b^0 & \Sigma_b^- \end{array} \right), \hspace{1ex} (B^{*ij})_\mu=\left( \begin{array}{cc} \Sigma_b^{*+} & \frac{1}{\sqrt{2}}\Sigma_b^{*0} \\ \frac{1}{\sqrt{2}}\Sigma_b^{*0} & \Sigma_b^{*-} \end{array} \right)_{\hspace{-1ex}\mu}. \\
\label{eq:Sfield}
\end{eqnarray}
On the other hand, the $s_l=0$ baryons ($J=1/2$) are antisymmetric in flavor and include only the $\Lambda_b$ in the $SU(2)$ case:
\begin{equation}
(T^{ij})=\frac{1}{\sqrt{2}}\left( \begin{array}{cc} 0 & \Lambda_b \\ -\Lambda_b & 0 \end{array} \right).  \label{eq:Tfield}
\end{equation}
The leading-order HH$\chi$PT Lagrangian, describing the interactions of the fields (\ref{eq:Hfield}),  (\ref{eq:Sfield}), and (\ref{eq:Tfield}) with pions,
is given by
\begin{eqnarray} 
\nonumber \mathcal{L} &=& (\:\:\mathrm{kinetic\hspace{1ex}terms}\:\:)\: +\: g_1\: \mathrm{tr}_{\mathrm{D}}\left[ \overline{H}_i (\mathscr{A}^{\mu})^i_{\:\:j} \gamma_\mu \gamma_5 H^j \right] \\
&&   -\: i \: g_2 \: \epsilon_{\mu\nu\sigma\lambda} \overline{S}^\mu_{ki} v^\nu (\mathscr{A}^\sigma)^i_{\:\:j} (S^\lambda)^{jk}\: +\: \sqrt{2} \: g_3 \left[ \overline{S}^\mu_{ki} (\mathscr{A}_\mu)^i_{\:\:j} T^{jk} + \overline{T}_{ki} (\mathscr{A}^\mu)^i_{\:\:j} S_\mu^{jk} \right]. \label{eq:HHchPTL}
\end{eqnarray}
In the terms with $g_1$, $g_2$, and $g_3$, the pion field $\xi=\sqrt{\Sigma}=\exp(i\Phi/f)$ appears through
\begin{equation}
\mathscr{A}^\mu = \frac i2 \left( \xi^\dag \partial^\mu \xi - \xi \partial^\mu \xi^\dag \right) = -\frac 1f \partial^\mu \Phi + ...\, ,
\end{equation}
which is an axial-vector field.

\section{Axial current matrix elements}

\begin{figure}[t]
  \centering
\includegraphics[width=0.35\columnwidth]{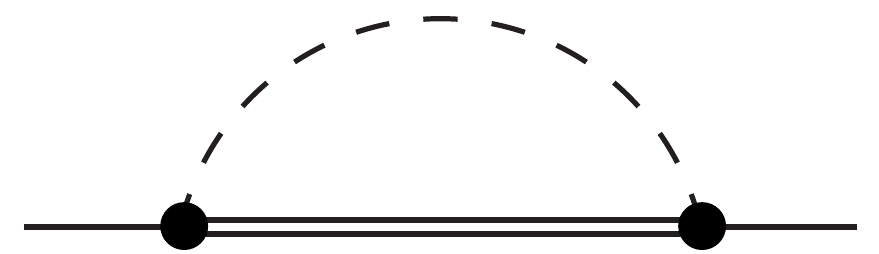}
\hspace*{1mm}
\includegraphics[width=0.225\columnwidth]{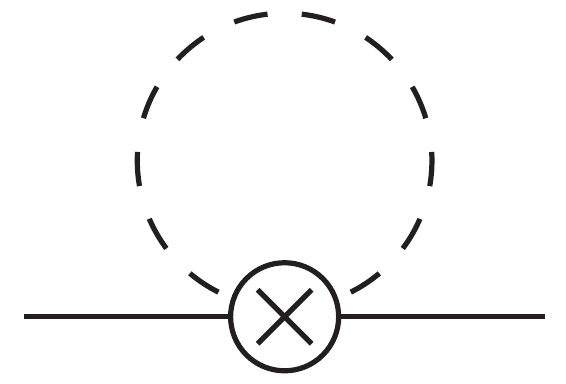}
\hspace*{1mm}
\includegraphics[width=0.35\columnwidth]{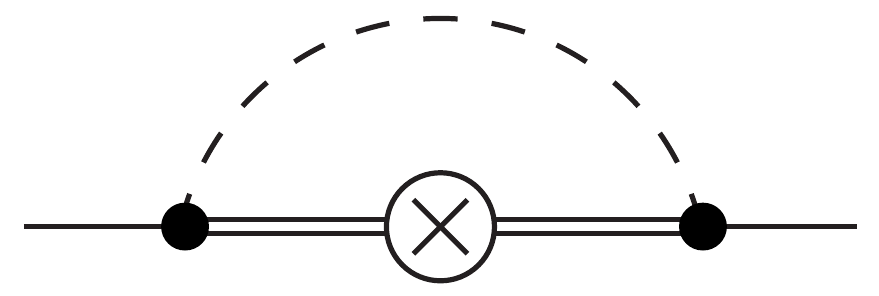} 
\\
(a)\hspace*{0.28\linewidth}(b)\hspace*{0.28\linewidth}(c)
  \caption{\label{fig:loops} One-loop diagrams contributing to the matrix elements of the axial current in HH$\chi$PT: (a) wavefunction renormalization diagram, (b) tadpole diagram, (c) sunset diagram \cite{Detmold:2011rb}. }
\end{figure}

To determine the axial couplings $g_i$ that appear in the chiral Lagrangian (\ref{eq:HHchPTL}) from QCD, one
can calculate suitable hadronic observables in both HH$\chi$PT and lattice QCD. The expressions derived
from HH$\chi$PT are then fitted to the lattice data, and in these fits the axial couplings are parameters.
The simplest observables that are sensitive to $g_i$ are the zero-momentum matrix elements of the axial current
between heavy-hadron states. In QCD, the isovector axial current is given by the quark current
\begin{equation}
 A_\mu^{a(\mathrm{QCD})} = \bar q \frac{\tau^a}{2} \gamma_\mu\gamma_5 q.
\end{equation}
The corresponding hadronic current in HH$\chi$PT can be obtained from the Lagrangian (\ref{eq:HHchPTL}) using the Noether procedure. At leading order,
the relevant part of the current that contributes to the matrix elements reads
\begin{eqnarray}
\nonumber A^{a(\chi \mathrm{PT},\mathrm{LO})}_\mu &=& \: {g_1} \: \mathrm{tr}_{\mathrm{D}} \left[ \ol{H}_i (\tau^a_{\xi+} )^i_{\:\:j} \gamma_\mu\gamma_5 H^j \right] - i {g_2} \: \varepsilon_{\mu\nu\sigma\lambda} \ol{S}^\nu_{ki}  v^\sigma ( \tau^a_{\xi+}  )^i_{\:\:j}  (S^\lambda)^{jk}\\
&& + \: \sqrt{2}\:{g_3} \left[ (\ol{S}_\mu)_{ki}  ( \tau^a_{\xi+}  )^i_{\:\:j}  T^{jk} + \ol{T}_{ki}  ( \tau^a_{\xi+}  )^i_{\:\:j}  (S_\mu)^{jk} \right], \label{eq:chPTaxialcurrent}
\end{eqnarray}
with $\tau^a_{\xi+} = \frac 12 \left( \xi^\dag \tau^a \xi + \xi \tau^a \xi^\dag \right).$ One finds the following matrix elements
for $A_\mu=A^{1}_\mu - i\:A^{2}_\mu$,
\begin{eqnarray}
  \nonumber  \langle P^{*d}  | A_\mu | P^u  \rangle &=& -2\:(g_1)_{\rm eff} \:\: \varepsilon^*_\mu, \\
  \nonumber  \langle S^{dd} | A_\mu | S^{du} \rangle &=& -(i/\sqrt{2})\:(g_2)_{\rm eff} \:\:v^\sigma
              \: \epsilon_{\sigma \mu \nu\rho}\: \overline{U}^\nu U^\rho, \\
  \langle S^{dd} | A_\mu | T^{du} \rangle &=& -(g_3)_{\rm eff} \:\:  \overline{U}_\mu \: \mathcal{U},
  \label{eq:matrixelts}
\end{eqnarray}
where $\varepsilon^\mu$ is the polarization vector of the $P^{*d}$ meson and
$\mathcal{U}$ is the Dirac spinor of the $T^{du}$ baryon. For the $s_l=1$ baryons,
we work directly with external $S^{ij}$ states (which contain the degrees of freedom of both $J=1/2$ and $J=3/2$)
and the $U^\mu$'s are the corresponding generalized spinors \cite{g123long}. At leading order in the chiral expansion,
the ``effective axial couplings'' in (\ref{eq:matrixelts}) are equal to the axial couplings in the Lagrangian,
$(g_i)_{\rm eff}\big|_{\rm LO}=g_i.$
At next-to-leading order, the matrix elements receive corrections from pion loops (Fig.~\ref{fig:loops})
and analytic counterterms. The NLO expressions for $(g_i)_{\rm eff}$, both in the unquenched and the partially quenched theories,
and for a finite volume, can be found in Ref.~\cite{Detmold:2011rb}.

\section{Lattice calculation}

To calculate the matrix elements (\ref{eq:matrixelts}) in lattice QCD, we use the following interpolating fields for the heavy hadrons,
\begin{eqnarray}
\nonumber P^i &=&  (\gamma_5)_{\alpha\beta} \: \overline{Q}_{a\alpha} \: \tilde{q}^i_{a\beta}, \hspace{10.25ex}
P^{*i}_\mu \:=\:  (\gamma_\mu)_{\alpha\beta} \: \overline{Q}_{a\alpha} \: \tilde{q}^i_{a\beta}, \\
 S^{ij}_{\mu\:\alpha} &=&  \epsilon_{abc}\:(C\gamma_\mu)_{\beta\gamma}\:\tilde{q}^i_{a\beta}\:\tilde{q}^j_{b\gamma}\: Q_{c\alpha}, \hspace{2ex} 
T^{ij}_\alpha \:=\: \epsilon_{abc}\:(C\gamma_5)_{\beta\gamma}\:\tilde{q}^i_{a\beta}\:\tilde{q}^j_{b\gamma}\: Q_{c\alpha}, \label{eq:interpolatingfields}
\end{eqnarray}
where $\tilde{q}^i$ denotes a smeared light-quark field of flavor $i$, and $Q$ denotes the static heavy quark field (here we set $v=(1,0,0,0)$).
Just as in HH$\chi$PT, the interpolating field $S^{ij}_{\mu\:\alpha}$ couples to both the $J=1/2$ and $J=3/2$ baryon states with $s_l=1$.
We use the domain-wall action \cite{Kaplan:1992bt} for the light quarks,
and the Eichten-Hill action \cite{Eichten:1989kb} with HYP-smeared temporal gauge links \cite{DellaMorte:2003mn}
for the heavy quarks. To optimize the signals and analyze heavy-quark discretization effects, we generated data for
$n_{\rm HYP}=1,2,3,5,10$ levels of HYP smearing. The final results for the axial couplings are based on data with $n_{\rm HYP}=1,2,3$ only.
The calculations are performed with the local 4-dimensional axial current, given by
\begin{equation}
A_\mu = Z_A\:\overline{d}_{a\alpha} (\gamma_\mu \gamma_5)_{\alpha\beta} u_{a\beta}, \label{eq:latticeaxialcurrent2}
\end{equation}
where $Z_A$ is determined nonperturbatively \cite{Aoki:2010dy}. We compute the following ratios of three-point and two-point functions,
\begin{eqnarray}
\nonumber R_1(t,\: t') &=& -\frac13 \frac{\sum_{\mu=1}^3 \langle \: P^{*d\:\mu}(t) \:A^\mu(t') \: P_u^{\dag}(0) \: \rangle}{\langle \: P^u(t) \: P_u^{\dag}(0) \: \rangle},  \\
\nonumber R_2(t,\: t') &=& i\:\:\frac{ \sum_{\mu,\nu,\rho=1}^3\epsilon_{0\mu\nu\rho} \: \langle \:S^{dd\:\mu}(t) \:A^\nu(t') \: \overline{S}_{du}^{\rho}(0) \: \rangle }{\sum_{\mu=1}^3 \langle \:S^{dd\:\mu}(t) \: \overline{S}_{dd}^{\mu}(0) \: \rangle },  \\
 R_3(t,\: t') &=& \left[\frac13\frac{\sum_{\mu,\nu=1}^3 \langle \:S^{dd\:\mu}(t) \:A^{\mu}(t') \: \overline{T}_{du}(0) \: \rangle \langle \:T^{du}(t) \:A^{\nu\dag}(t') \: \overline{S}_{dd}^{\nu}(0) \: \rangle }{ \sum_{\mu=1}^3 \langle \:S^{dd\:\mu}(t) \: \overline{S}_{dd}^{\mu}(0) \: \rangle \:\langle \:T^{du}(t) \: \overline{T}_{du}(0) \: \rangle  }\right]^{1/2}, \label{eq:ratios}
\end{eqnarray}
where the source and sink hadron interpolating fields are placed at a common spatial point $\mathbf{x}$ because of the static heavy quark,
and we write $A^\mu(t')=\sum_{\mathbf{x'}}A^\mu(t',\mathbf{x'})$.
In Eq.~(\ref{eq:ratios}) we also removed the free spinor indices which trivially originate from the static heavy-quark propagator.
By inserting complete sets of states into (\ref{eq:ratios}), one can show that $R_i(t,\:t/2) = (g_i)_{\rm eff}+...\,$,
where the dots indicate contributions from excited states that decay exponentially with $t$ \cite{g123long}.

\begin{figure}[t]
\begin{center}
  \includegraphics[height=0.385\linewidth]{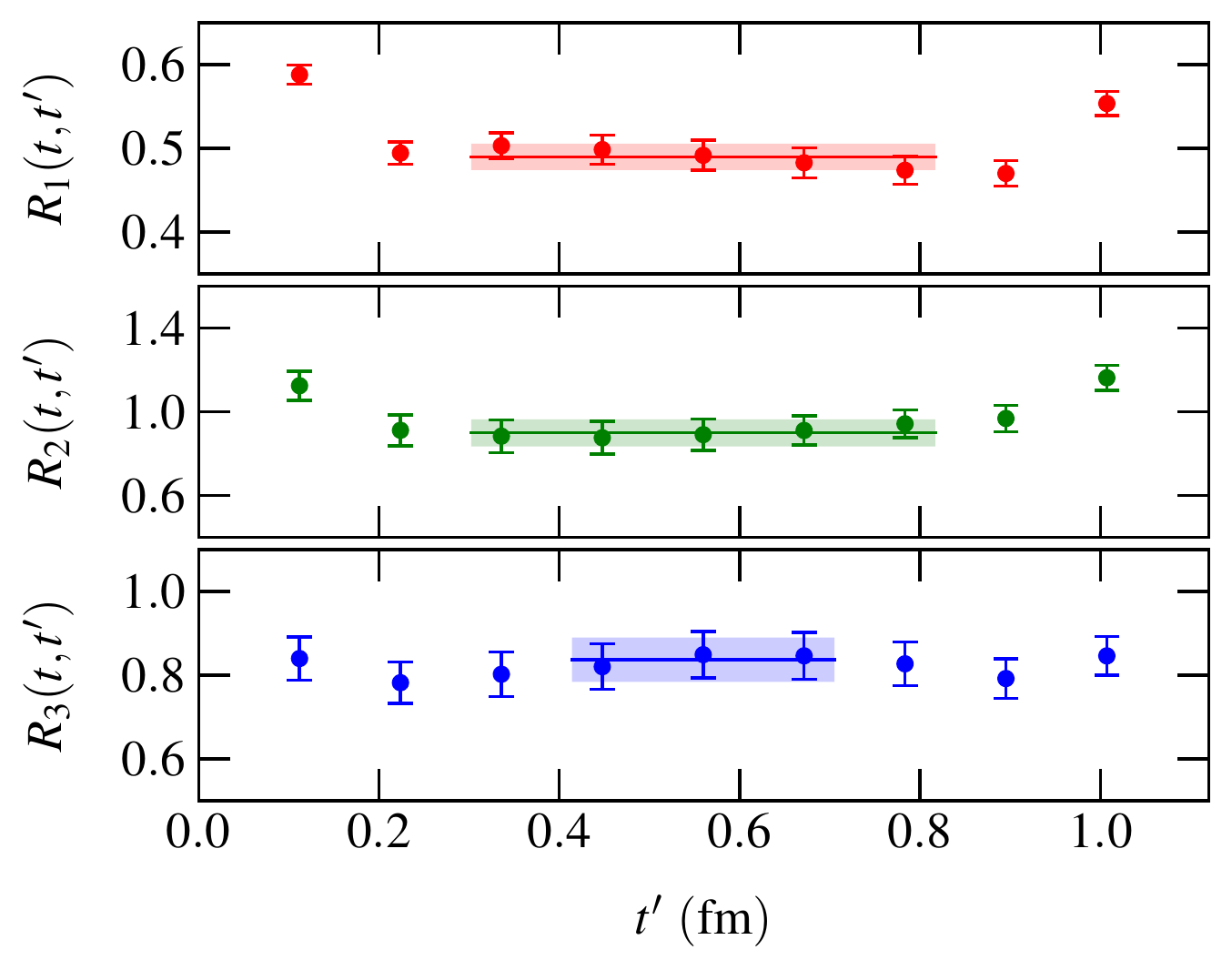} \hfill \includegraphics[height=0.385\linewidth]{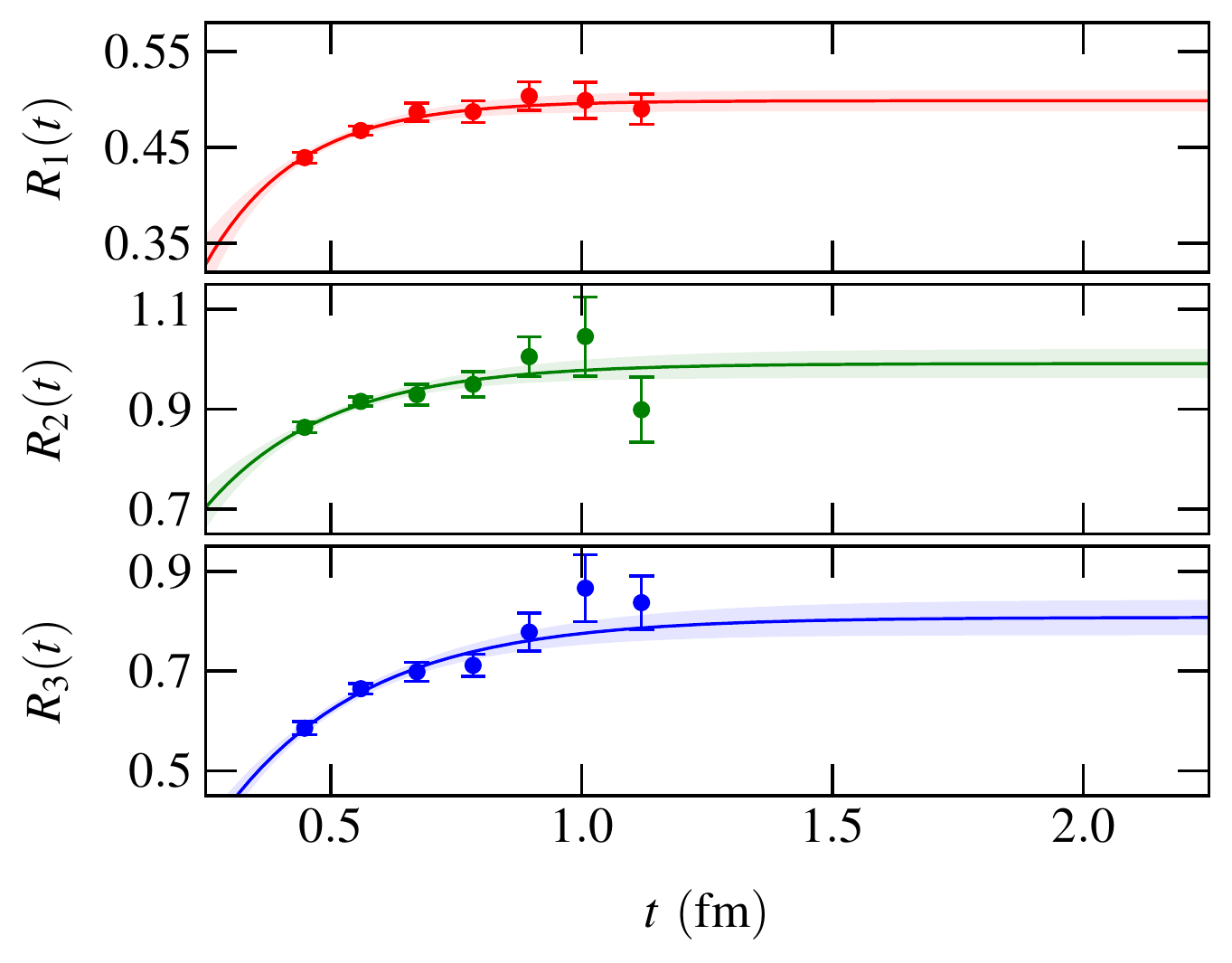}
\vspace{-1ex}
  \caption{\label{fig:ratios}Left panel: ratios $R_i(t,\:t')$ for the source-sink separation $t/a=10$, along with extracted values $R_i(t)$ (shaded regions).
        Right panel: extrapolation of $R_i(t)$ to infinite source-sink separation. All data shown here are for $a=0.112$ fm, $am_{u,d}^{(\mathrm{val})}=0.002$, $n_{\rm HYP}=3$.}
\end{center}
\vspace{-3ex}
\end{figure}

Our calculations use RBC/UKQCD gauge field configurations with 2+1 dynamical
quark flavors \cite{Aoki:2010dy}. The main parameters
of the ensembles and the domain-wall propagators we computed on them are given in Table \ref{tab:params}. For the three-point
functions we use pairs of light-quark propagators with sources at a common spatial point $\mathbf{x}$ and separated by $t/a$
steps in the time direction. As can be seen in the table, we have data for multiple values of $t/a$.
Numerical examples for the ratios (\ref{eq:ratios}) are shown in Fig.~\ref{fig:ratios} (left).
Equivalently to using $R_i(t,\:t/2)$, we average $R_i(t,\:t')$ over $t'$ in the central plateau regions, and we denote these averages as $R_i(t)$.
We then perform fits of the form $R_i(t)=(g_i)_{\rm eff}-A_i\:e^{-\delta_i \: t}$, as shown in Fig.~\ref{fig:ratios} (right). A detailed
discussion can be found in Ref.~\cite{g123long}. These fits provide
the effective axial couplings $(g_i)_{\rm eff}(a, m_\pi^{(\mathrm{vv})}, m_\pi^{(\mathrm{vs})}, n_{\rm HYP})$
for all combinations of the lattice spacing, the pion masses, and the heavy-quark smearing parameter $n_{\rm HYP}$.
\begin{table}[t]
\begin{center}
\begin{tabular}{ccccccl}
\hline\hline
$L^3\times T$ & $am_{u/d}^{(\mathrm{sea})}$ & $am_{u/d}^{(\mathrm{val})}$     & $a$ (fm)                    & $m_\pi^{(\mathrm{vv})}$ (MeV) & $m_\pi^{(\mathrm{vs})}$ (MeV) & values of $t/a$  \\
\hline
$24^3\times64$ & $0.005$ & $0.005$  & $0.1119(17)$ & 336(5) & 336(5) & 4, 5, 6, 7, 8, 9, 10  \\
$24^3\times64$ & $0.005$ & $0.002$  & $0.1119(17)$ & 270(4) & 304(5) & 4, 5, 6, 7, 8, 9, 10  \\
$24^3\times64$ & $0.005$ & $0.001$  & $0.1119(17)$ & 245(4) & 294(5) & 4, 5, 6, 7, 8, 9, 10  \\
$32^3\times64$ & $0.006$ & $0.006$  & $0.0848(17)$ & 352(7) & 352(7) & 13                     \\
$32^3\times64$ & $0.004$ & $0.004$  & $0.0849(12)$ & 295(4) & 295(4) & 6, 9, 12               \\
$32^3\times64$ & $0.004$ & $0.002$  & $0.0849(12)$ & 227(3) & 263(4) & 6, 9, 12               \\
\hline\hline
\end{tabular}
\caption{\label{tab:params}Lattice parameters. $m_\pi^{(\mathrm{vv})}$ and $m_\pi^{(\mathrm{vs})}$ denote the valence-valence and valence-sea pion masses.}
\end{center}
\vspace{-2ex}
\end{table}
We fit the data for $(g_i)_{\rm eff}(a, m_\pi^{(\mathrm{vv})}, m_\pi^{(\mathrm{vs})}, n_{\rm HYP})$ with
\begin{eqnarray}
\nonumber (g_1)_{\rm eff} &=& g_1\Big[1 + f_1(g_1,\:m_\pi^{(\mathrm{vv})},\: m_\pi^{(\mathrm{vs})},\: L) + c_1^{(\mathrm{vv})}\: [m_\pi^{(\mathrm{vv})}]^2 + c_1^{(\mathrm{vs})}\: [m_\pi^{(\mathrm{vs})}]^2 + d_{1,\:n_{\rm HYP}}\:\:a^2 \Big], \\
 (g_{i})_{\rm eff}\Big|_{i=2,\:3}\hspace{-1ex} &=& g_{i}\Big[1 + f_{i}(g_2,\: g_3,\: m_\pi^{(\mathrm{vv})},\: m_\pi^{(\mathrm{vs})},\: \Delta,\: L) + c_{i}^{(\mathrm{vv})}\: [m_\pi^{(\mathrm{vv})}]^2 + c_{i}^{(\mathrm{vs})}\: [m_\pi^{(\mathrm{vs})}]^2 + d_{i,\:n_{\rm HYP}}\:\:a^2 \Big],\phantom{XXX} \label{eq:geffNLO}
\end{eqnarray}
where the functions $f_i$ are the NLO loop contributions from $SU(4|2)$ HH$\chi$PT, including the effects of the finite lattice size \cite{Detmold:2011bp}. The terms with
coefficients $c_i^{(\mathrm{vv})}$ and $c_i^{(\mathrm{vs})}$ are analytic NLO counterterms that cancel the renormalization-scale-dependence of $f_i$, and the terms
with coefficients $d_{i,\:n_{\rm HYP}}$ describe the leading effects of the non-zero lattice spacing. The functions $f_{2,3}$ also depend on the $S-T$ mass splitting
$\Delta$, which is included in the kinetic terms of Eq.~(\ref{eq:HHchPTL}). We set $\Delta=200$ MeV, consistent with the
$\Sigma_b^{(*)}-\Lambda_b$ splitting from experiment and with our lattice data.

To study the effect of the HYP smearing in the heavy-quark action on the scaling behavior, we performed initial fits
that included all values of $n_{\rm HYP}$, and then successively removed the data with the largest values of $n_{\rm HYP}$. After excluding
$n_{\rm HYP}=5,10$, the fits were stable and had good $Q$-values. Our final results for the axial couplings are
\begin{eqnarray}
  g_1&=&0.449   \pm 0.047_{\:\rm stat}   \pm 0.019_{\:\rm syst}, \nonumber \\
  g_2&=&0.84\nb \pm 0.20_{\:\rm stat}\nb \pm 0.04_{\:\rm syst},  \nonumber \\
  g_3&=&0.71\nb \pm 0.12_{\:\rm stat}\nb \pm 0.04_{\:\rm syst}. \label{eq:finalresults}
\end{eqnarray}
The estimates of the systematic uncertainties in (\ref{eq:finalresults}) include the effects of the following: NNLO terms in the fits
to the $a$- and $m_\pi$-dependence (3.6\%, 2.8\%, 3.7\% for $g_1$, $g_2$, $g_3$, respectively),
the above-physical value of the sea-strange-quark mass (1.5\%), and higher excited states in $R_i(t)$
(1.7\%, 2.8\%, 4.9\%). The details of the analysis can be found in Ref.~\cite{g123long}.

\begin{figure}[t]
\begin{center}
  \includegraphics[height=0.48\linewidth]{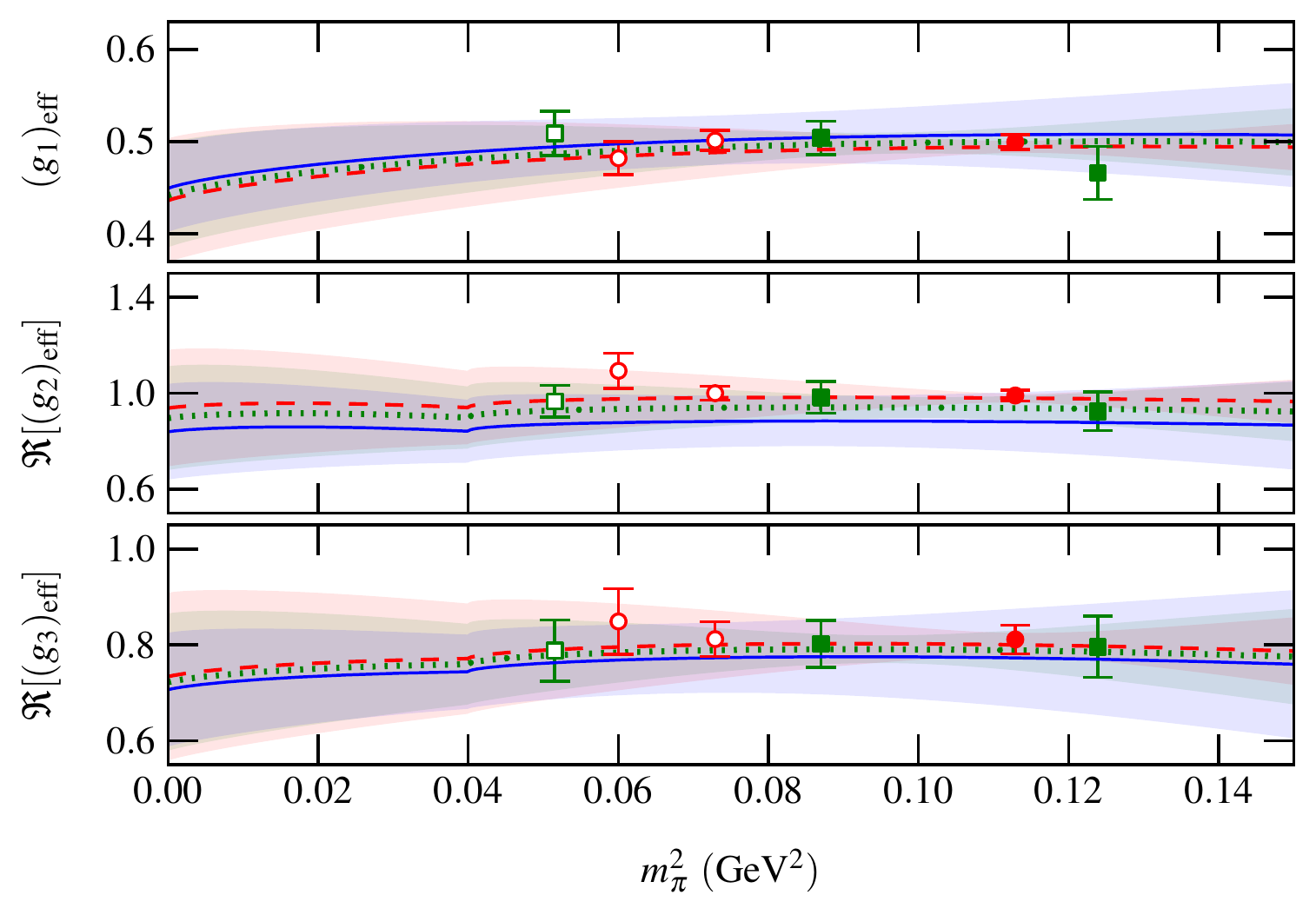}
  \caption{\label{fig:madep} Fits of $(g_i)_{\rm eff}$ using Eq.~(\protect\ref{eq:geffNLO}). The plot shows the
  fitted functions, evaluated at $m_\pi^{(\mathrm{vv})}=m_\pi^{(\mathrm{vs})}=m_\pi$ and
  in infinite volume, for $n_{\rm HYP}=3$. The baryonic matrix elements $(g_{2,3})_{\rm eff}$ develop small imaginary parts
below the $S \to T \pi$ decay threshold at $m_\pi=\Delta$, and only the real parts 
are shown here.
  The dashed line corresponds to $a=0.112$ fm, the dotted line to $a=0.085$ fm, and the solid line
  to the continuum limit. The $\pm 1 \sigma$ regions are shaded. The data points (circles: $a=0.112$ fm, squares:
  $a=0.085$ fm) have been shifted to infinite volume for this plot, and the partially quenched data ($m_\pi^{(\mathrm{vv})}<m_\pi^{(\mathrm{vs})}$)
  are included using open symbols at $m_\pi=m_\pi^{(\mathrm{vv})}$, even though the fit functions have slightly different values for these points.
 }
\end{center}
\end{figure}

Figure \ref{fig:madep} shows the pion-mass dependence of the fitted functions $(g_i)_{\rm eff}$.
The counterterm parameters $c_i^{(\mathrm{vv})}$ and $c_i^{(\mathrm{vs})}$ resulting from the fits are natural-sized (for $\mu=4\pi f_\pi$),
and the NLO contributions are significantly smaller than the LO contributions. We conclude that the $SU(4|2)$ chiral expansion of $(g_i)_{\rm eff}$
convergences well for the pion masses used here.

\section{Summary}

We have calculated the heavy-hadron axial couplings using lattice QCD, including for the first time the baryonic couplings $g_2$ and $g_3$
in addition to the mesonic coupling $g_1$.
The analysis is based on data for the axial-current matrix elements at low pion masses, a large volume, and two different lattice spacings.
We extracted $g_{1,2,3}$ from this data by performing chiral fits with the full NLO expressions from HH$\chi$PT \cite{Detmold:2011rb}.
As a consequence, the systematic uncertainties in our results (\ref{eq:finalresults})
are much smaller than the statistical uncertainties.
The numerical values of $g_{1,2,3}$ can be used to constrain
chiral fits of lattice QCD data for a wide range of heavy-light meson and baryon observables.
Furthermore, our results for the axial couplings allow the direct calculation
of certain observables in HH$\chi$PT, in particular the strong decay widths of heavy baryons \cite{Detmold:2011bp, g123long}.

\vspace{4ex}

\textbf{Acknowledgments:}
WD is supported in part by JSA, LLC under DOE contract No.~DE-AC05-06OR-23177 and by the Jeffress Memorial Trust,
J-968. WD and SM are supported by DOE OJI Award D{E}-S{C}000-1784 and DOE grant
DE-FG02-04ER41302. CJDL is supported by NSC grant number 99-2112-M-009-004-MY3.
We acknowledge the hospitality of Academia Sinica Taipei and NCTS Taiwan. This research
made use of computational resources provided by NERSC and the NSF Teragrid.

\end{document}